\documentclass[12pt]{article}
\usepackage{graphicx}
\usepackage{amsmath}

\title{ \bf{Endpoints of invariant mass distribution in SUSY particle decays
into massive particles}}

\author{\textbf{J. DiBello$^{1}$ , R. Konoplich$^{1,2}$, N. Lavini$^{1}$, T. St.Laurent$^{1}$ }\\
        \normalsize$^{1}$Department of Physics, Manhattan College, Riverdale, New York, NY, 10471\\
        \normalsize$^{2}$Department of Physics, New York University, New York, NY 10003}

\begin{document}

\maketitle

\begin{abstract}
Kinematic limits on an invariant mass distribution of bc-pairs
for a three-step decay chain $A \rightarrow bB \rightarrow bcC$ involving all 
massive particles are found. It is shown that an application of these limits to
a stop quark production at the LHC could reduce significantly 
Standard Model background contribution.

\end{abstract}

\newpage

It was shown \cite{baer}-\cite{bar2} that the endpoint method could be very useful
in SUSY particle mass reconstruction to find relations between masses of
SUSY particles involved in a decay chain and to determine their masses.
This method allows mass reconstruction without relying on a specific SUSY model.
In particular the endpoint method can be applied to the decay chain 

\begin{equation}
A \to b B \to b c C .
\label{chain} 
\end{equation}

\noindent
where particles A, B, C are invisible but particles b and c can be 
either detected or reconstructed and are considered as visible.

\begin{figure}[h]
  \centering
  \includegraphics[width=0.7\textwidth]{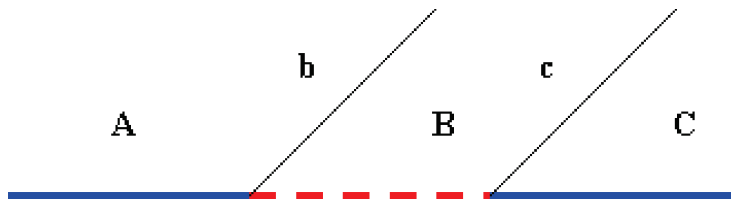} 
  \caption[Short caption.]{A cascade decay chain.}
\label{fig:chain}
\end{figure}

This decay chain (\ref{chain}) is shown in Fig. \ref{fig:chain}

In literature, kinematic limits on an invariant mass distribution of bc-pairs
in decay (\ref{chain}) over a variable $q^2=(p_b+p_c)^2$ often are given for a case
when at least one of visible particles is massless. However, in some cases
both particles b and c can have non-negligible mass. For example, this is the 
case when a gluino
decays into a stop quark and top quark \cite{noj1},\cite{noj2}. 
We derived these
kinematic limits for the case of all massive particles in process (\ref{chain})
and we found

\begin{equation}
q=\sqrt{[-R \pm \sqrt{R^2-4QS}]/2Q} ,
\label{limits}
\end{equation}
\begin{equation}
Q=M_B^2 ,
\end{equation}
\begin{equation}
R=(m_b^2-M_A^2-M_B^2)(m_b^2+m_c^2)+(m_b^2-M_A^2+M_B^2)(M_B^2-m_b^2-M_C^2) ,
\end{equation}
\begin{equation}
S=M_A^2(m_b^2-m_c^2)^2+(M_A^2-M_C^2)(m_b^2(M_B^2-m_b^2-M_C^2)+m_c^2(M_A^2+m_b^2-M_B^2)) ,
\end{equation}

\noindent
where an upper edge corresponds to the case when b and c particles are moving in opposite
directions in the rest frame of particle A and a lower edge corresponds to the 
case when b and c particles are moving in the same direction. A nonzero
lower limit is a consequence of nonzero masses of particles. Note that a similar formula
can be found in \cite{bartl} or obtained from Eqs.(E.9),(E.10) of \cite{lester_thes}.

In order to demonstrate the possibility of using these kinematic edges for a background 
suppression we consider a stop quark production in gluino decay 

\begin{equation}
\tilde{g} \to \tilde{t_1} t \to \tilde{\chi}_{2}^{0} t t . 
\label{stop} 
\end{equation}

where $\tilde{\chi}_{2}^{0}$ decays into $\tilde{l_R} l \to \tilde{\chi}_{1}^{0} l l$.

The study of sleptons and squarks of third generation is of special interest.
Their masses can be very different than those of sparticles of the first and second 
generation, because of the effects of large Yukawa and soft couplings
in the renormalization group equations. Furthermore, they can show large mixing in
pairs $(\tilde{t}_{L}, \tilde{t}_{R}), (\tilde{b}_{L}, \tilde{b}_{R})$
and $(\tilde{\tau}_{L}, \tilde{\tau}_{R})$. 

For this study we choose the SU3 model point. The bulk point SU3 
is the official benchmark point of the ATLAS collaboration at the LHC 
and it is in agreement with
the recent precision WMAP data ~\cite{wmap}. 
This model point is described by the
set of mSUGRA parameters given in Table \ref{tab:paramth}.

\begin{table} [h]
\begin{center}
 \begin{tabular}{|c|c|c|c|c|c|}
   \hline
   Point & $m_{0}$ & $m_{1/2}$ & $A_{0}$ & $tan\beta$ & $\mu$ \\
   \hline\hline
   SU3 & 100 GeV & 300 GeV & -300 GeV & 6 &  $>$ 0 \\
   \hline
 \end{tabular}
\caption{mSUGRA parameters for the SU3 point.}\label{tab:paramth}
\end{center}
\end{table} 

The possibilities for a stop quark reconstruction in different 
decay chains and for different points in the MSSM parameter space 
were discussed, for example in \cite{gj_note},\cite{krstic}-\cite{mehdi}.  
Recent results on searches for stop quarks were published in \cite{cdf}.

Assumed theoretical masses of SUSY particles in the cascade (\ref{chain})  
and a cross section
generated by ISAJET 7.74 \cite{isajet} are given in Table \ref{tab:massth}.

\begin{table} [h]
\begin{center}
 \begin{tabular}{|c|c|c|c|c|c|c|}
   \hline
   Point & $m_{\tilde{g}}$ & $m_{\tilde{t_{1}}}$ & $m_{\tilde{\chi}_{2}^{0}}$ & $m_{\tilde{l}_{R}}$ & $m_{\tilde{\chi}_{1}^{0}}$ & $\sigma [\rm pb]$ \\
   \hline\hline
   SU3 & 720.16 & 440.26 &  223.27 & 151.46 & 118.83 & 19\\
   \hline
 \end{tabular}
\caption{The assumed  theoretical masses of sparticles BR and the production cross section $\sigma $ 
at the SU3 point. 
Masses are given in GeV.}\label{tab:massth}
\end{center}
\end{table}  

A branching ratio for the gluino decay chain (\ref{stop}) at the SU3 point is

\begin{equation}
\nonumber
\tilde{g} \stackrel{25.2 \%}{\longrightarrow} \tilde{t}_{1} \stackrel{11.5 \%}{\longrightarrow} \tilde{\chi}_{2}^{0} 
\stackrel{11.4 \%}{\longrightarrow} \tilde{l}_{R} \stackrel{100 \%}{\longrightarrow} \tilde{\chi}_{1}^{0} ~~\Rightarrow ~~0.33 \% . 
\end{equation}

\noindent 
Stop quark $\tilde{t}_{1}$ is a mixture of the $\tilde{t}_{L}$ and $\tilde{t}_{R}$ states. 
At the SU3 point,
$\tilde{t}_{1}$ is the lightest supersymmetric quark because of the renormalization 
group equation 
running effect and because $\tilde{t}_{1}$ mass is related with the Higgs mass through
radiation corrections.

Monte Carlo simulations of SUSY production at model points were performed by the HERWIG 6.510
event generator  ~\cite{herwig}. The produced events were passed through the AcerDET 
detector simulation ~\cite{atlfast}, which parametrized the response of a generic
detector (LHC detector descriptions can be found in \cite{det_atlas}, \cite{det_cms}).  
Samples of 400k  SUSY events were used. 
This approximately  corresponds  to $20~\rm fb^{-1}$ of integrated luminosity
for the SUSY SU3 point production cross section of 19 pb at 14 TeV.

In order to isolate the chain (\ref{stop}) and to suppress the backgrounds 
the following selection cuts were applied:

$\bullet$ two isolated opposite-sign same-flavor (OSSF) leptons (not tau leptons)
satisfying transverse momentum cuts $p_{T}(l^{\pm}) > 20~GeV$ and 
$p_{T}(l^{\mp}) >  10~GeV$

$\bullet$ two b-tagged jets with $p_{T} > 50~GeV$; 

$\bullet$ at least three jets, the hardest satisfying 
$p_{T1} > 150~GeV$, $p_{T2} > 100~GeV$, $p_{T3} > 50~GeV$;  

$\bullet$ total number of jets (including b-tagged jets) $N_{jet} \ge 7$
satisfying $p_{T} > 10~GeV$;

$\bullet$ no $\tau$-tagged jets 

$\bullet$ $M_{eff} > 600~GeV$ and $E_{T}^{miss} > 0.2M_{eff}$, where
$E_{T}^{miss}$ is the missing transverse energy and $M_{eff}$ is the scalar
sum of the missing transverse energy and the transverse momenta of the four
hardest jets; 

$\bullet$ lepton invariant mass $50~GeV < M_{ll} < 105~GeV$.

Top quarks appear both in the signal and in backgrounds.
The most important backgrounds for the process (\ref{stop}) are 
Standard Model $t \bar t$ background and SUSY background
when $t \bar t$ quarks are produced in processes involving SUSY particles but
in decay chains different than that in process (\ref{stop}).

After selection cuts were applied, we found 26 signal events corresponding
to process (\ref{stop}), 63 SUSY background events containing 
$t \bar t$ quarks and 155 Standard Model $t \bar t$ events.

\begin{figure}[htb!]
  \centering
  \includegraphics[width=0.6\textwidth]{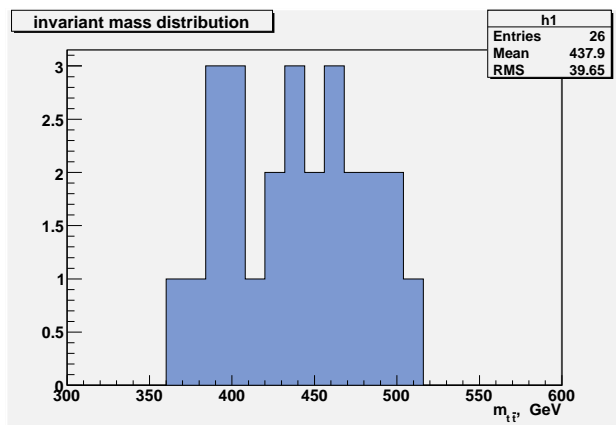} 
  \caption[Short caption.]{$t \bar t$ invariant mass distribution for the signal.} 
  \label{fig_signal}
\end{figure}

Fig. \ref{fig_signal}  shows the $t \bar t$ invariant mass 
distribution for 26 signal events remaining
after application of the kinematic cuts. At this step, truth information
for momenta and energies of $t \bar t$ quarks 
was used. 
One can see that in this case only two events are outside these kinematic limits 
which, for process (\ref{stop}), are $q_{min}$ = 375.1 GeV and $q_{max}$ = 496.8 GeV, 
respectively.

\begin{figure}[htb!]
  \centering
  \includegraphics[width=0.6\textwidth]{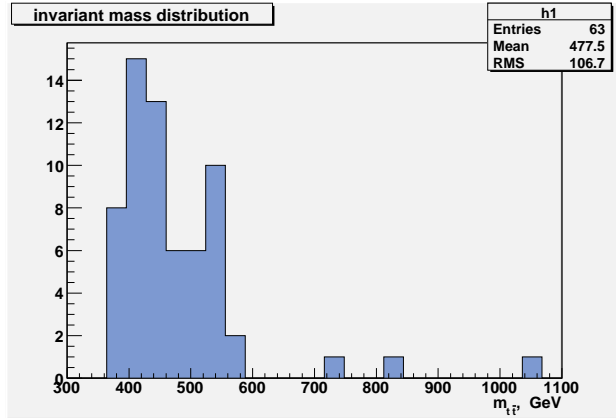} 
  \caption[Short caption.]{$t \bar t$ invariant mass distribution for SUSY background.} 
  \label{fig_susybkg}
\end{figure}

\begin{figure}[htb!]
  \centering
  \includegraphics[width=0.6\textwidth]{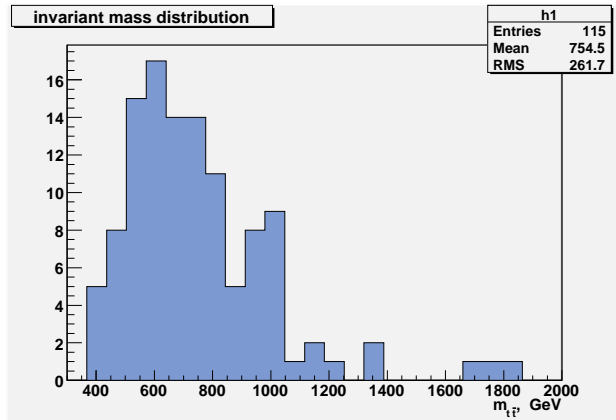} 
  \caption[Short caption.]{$t \bar t$ invariant mass distribution for Standard Model 
  $t\bar{t}$ background.} 
  \label{fig_ttbar}
\end{figure}

Figs. \ref{fig_susybkg}, \ref{fig_ttbar} show 
the $t \bar t$ invariant mass distribution for SUSY background and 
Standard Model $t \bar t$ events. In the last case, most of events are
outside kinematic limits (\ref{limits}).

To suppress Standard Model and SUSY backgrounds an additional cut on events
was applied by using upper and lower limits given by Eq.(\ref{limits}). 
Table \ref{tab:supress} shows the number of signal events, SUSY
background events and Standard Model $t \bar t$ events before and after kinematic cuts 
(\ref{limits}) were applied.

\begin{table} [htb!]
\begin{center}
 \begin{tabular}{|c|c|c|c|}
   \hline
   Total & Signal & SUSY backg. & $t \bar t$ backg.\\
   \hline\hline
   204/75 & 26/24 & 63/40 & 115/11\\
   \hline
 \end{tabular}
\caption{The number of signal and background events before and after 
application of kinematic cuts (\ref{limits}).}
\label{tab:supress}
\end{center}
\end{table} 

It follows from Table \ref{tab:supress} that the number of events 
surviving selection cuts is reduced significantly after the application
of kinematic cuts (\ref{limits}) especially for Standard Model $t \bar t$ background
events.

These results show that the application of kinematic limits can be 
effective for background suppression in searches for supersymmetry
at the LHC.

This work has been supported in part by the National Science Foundation under grant PHY-0854724.

\end{document}